\pdfoutput=1
\documentclass[aps,prd,amsmath,floats,floatfix, twocolumn,
superscriptaddress,nofootinbib,showpacs]{revtex4-2}

\usepackage[T1]{fontenc}
\usepackage[utf8]{inputenc}
\usepackage{lmodern}
\usepackage[caption=false]{subfig}
\usepackage{verbatim}
\usepackage{multirow}
\usepackage[dvipsnames, usenames]{xcolor}
\usepackage{tikz}
\usetikzlibrary{decorations.pathmorphing}
\usetikzlibrary{decorations.markings}

\definecolor{linkcolor}{rgb}{0.0,0.3,0.5}
\usepackage[hypertexnames=false, unicode, colorlinks=true, linkcolor=linkcolor,
citecolor=linkcolor, filecolor=linkcolor,urlcolor=linkcolor,
pdfusetitle]{hyperref}

\usepackage[all]{hypcap}
\usepackage{graphicx}
\usepackage{xspace}
\usepackage{amssymb}
\usepackage[normalem]{ulem} 
\usepackage{bm} 
\usepackage{enumitem,amssymb}

\usepackage{microtype}
\usepackage{orcidlink}

\usepackage[english]{babel}
\usepackage{blindtext}

\graphicspath{%
  {figs/}%
}

\DeclareMathAlphabet{\mathpzc}{OT1}{pzc}{m}{it}

\newlist{todolist}{itemize}{2}
\setlist[todolist]{label=$\square$}

\begin{document}

\title{Starclashes: X-Ray Bursts from Relativistic Stellar Collisions near Massive Black Holes}

\author{Huan Yang}
\email{hyangdoa@tsinghua.edu.cn}
\affiliation{Department of Astronomy, Tsinghua University, Beijing 100084, China}

\author{Mengqi Yang}
\affiliation{Tsung-Dao Lee Institute, Shanghai Jiao Tong University, Shanghai, 201210, China}
\affiliation{State Key Laboratory of Dark Matter Physics, Key Laboratory for Laser Plasmas (MOE), School of Physics and Astronomy, Shanghai Jiao Tong University, Shanghai 200240, China}

\author{Houyi Sun}
\affiliation{Department of Astronomy, Tsinghua University, Beijing 100084, China}

\author{Hong-Xuan Jiang}
\affiliation{The Hong Kong Institute for Astronomy and Astrophysics, The University of Hong Kong, Pokfulam Road, Hong Kong, China}

\author{Xinyu Li}
\email{xinyuli@tsinghua.edu.cn}
\affiliation{Department of Astronomy, Tsinghua University, Beijing 100084, China}

\author{Ning Jiang}
\affiliation{Department of Astronomy, University of Science and Technology of China, Hefei, 230026, China}

\begin{abstract}

The star–disk collision interpretation of quasi-periodic eruptions suggests that some supermassive black holes host stellar companions at orbital separations of order hundreds of gravitational radii. Using time-dependent Fokker–Planck models of nuclear star clusters, we estimate that these companions undergo collisions with surrounding stars at rates of $\mathcal{O}(10^{-5}-10^{-4})\,\mathrm{yr}^{-1}$ per companion for black hole masses of $10^6-10^8 M_\odot$. With relative velocities of order $0.1c$, such starclashes generate energetic, shock-heated fireballs. Our three-dimensional radiation hydrodynamics simulations predict luminous X-ray bursts lasting one to several hours. Reprocessing of the fireball radiation by nearby gas clouds and interaction of the ejecta with the gaseous medium may produce optical and radio counterparts, respectively. Subsequent tidal disruption and accretion of the collision debris may power additional emission. These transients may provide an independent probe of stellar populations in the immediate vicinity of supermassive black holes.
\end{abstract}

\maketitle

\noindent{\it Introduction-}
Quasi-periodic eruptions (QPEs)~\cite{Miniutti2019Discovery} have motivated extensive investigations into their physical origin, with proposed mechanisms including disk instabilities~\cite{Pan2022QPE,Kaur2023QPE}, partial tidal disruptions~\cite{King2020QPE,Wang2022QPE}, and star--disk collisions~\cite{Xian2021QPE,LinialMetzger2023QPE}. 
The alternating long and short waiting times observed in GSN 069 arise naturally in the star–disk collision model when the stellar companion follows an eccentric orbit~\cite{Miniutti2023QPE,LinialMetzger2023QPE}.
Recent observations suggest that $9^{+9}_{-5}\%$ of optically selected tidal disruption events (TDEs) exhibit QPEs within five years~\cite{Chakraborty2025AT2022upj}, supporting the scenario in which a pre-existing stellar companion interacts with a transient TDE disk, as proposed for AT2019qiz~\cite{Nicholl2024AT2019qiz}. 

Fits to QPE timing data indicate characteristic orbital radii of order $10^2$ gravitational radii~\cite{Zhou2024QPEOrbits,Zhou2024QPEPopulation}. 
Independent evidence that stars can reach these radii comes from S301, whose highly eccentric orbit brings it within approximately $280\,GM_\bullet/c^2$ of Sgr~A* at pericenter~\cite{GRAVITY2026S301}.

One plausible origin for these close companions in QPE systems is inward migration through an accretion disk during an earlier active galactic nucleus (AGN) phase~\cite{PanYang2021,JiangPan2025Embers}. Disk migration can efficiently transport embedded stars toward the black hole, and, in some disk models, inward migration slows near $10^2$ gravitational radii~\cite{Zhou2024QPEOrbits}. Extended emission-line regions in three of five surveyed QPE hosts indicate a brighter past nucleus, consistent with this scenario~\cite{Wevers2024Hosts}. 
Stars deposited on such tightly bound orbits may persist after the AGN disk dissipates and thus remain largely unobservable until a subsequent dynamical interaction occurs. If the star–disk collision interpretation of QPEs is correct, what other observable signatures might arise from this population of close stellar companions?

\begin{figure*}[t]
  \centering
  \includegraphics[width=0.9\textwidth]{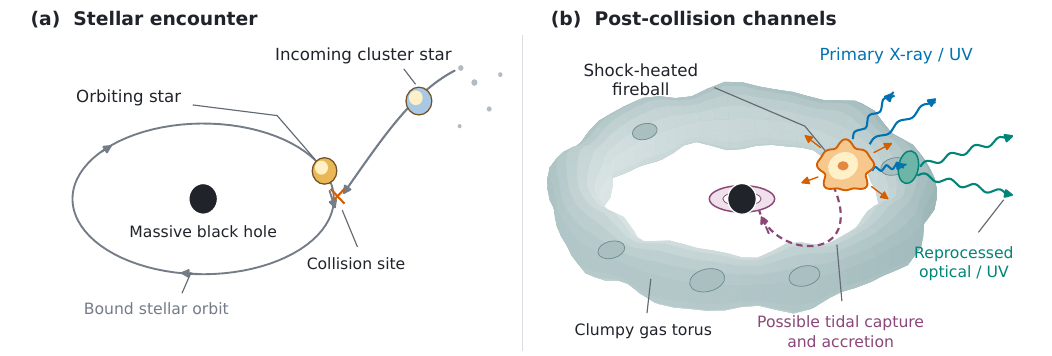}
  \caption{Stellar collisions near a massive black hole and possible
  electromagnetic counterparts. (a) A nuclear-cluster star approaches a close
  stellar companion on an intersecting trajectory. (b) The resulting
  shock-heated fireball emits primary X-ray/UV radiation. Nearby gas,
  illustrated as a clumpy torus, intercepts part of this radiation and
  reprocesses it into optical/UV emission.  There is a finite chance for the post-collision fireball(s) moving towards the massive black hole and producing a tidal disruption event. Bound debris may be tidally
  stretched and accreted.}
  \label{fig:collision-system}
\end{figure*}

Fast stellar collisions have long been considered a source of gas in dense galactic nuclei~\cite{SpitzerSaslaw1966} and a possible source of nuclear variability near supermassive black holes~\cite{Courvoisier1996Collisions}. Balberg, Sari, and Loeb predicted explosive encounters involving stars captured by tidal separation of binaries~\cite{Balberg2013Collisions}. 
Here we consider a different population of collision targets - the starclashes, as motivated by the star–disk collision interpretation of QPEs. A close stellar companion can be struck by stars scattered onto low-angular-momentum orbits in the surrounding nuclear star cluster, as illustrated in Fig.~\ref{fig:collision-system}.
Using a Fokker--Planck framework previously developed to model the rates of extreme mass-ratio inspirals ~\cite{PanYang2021}, we find collision rates of order $10^{-5}-10^{-4}\,{\rm yr}^{-1}$ per companion.

\begin{figure}
    \centering
    \includegraphics[width=0.95\linewidth]{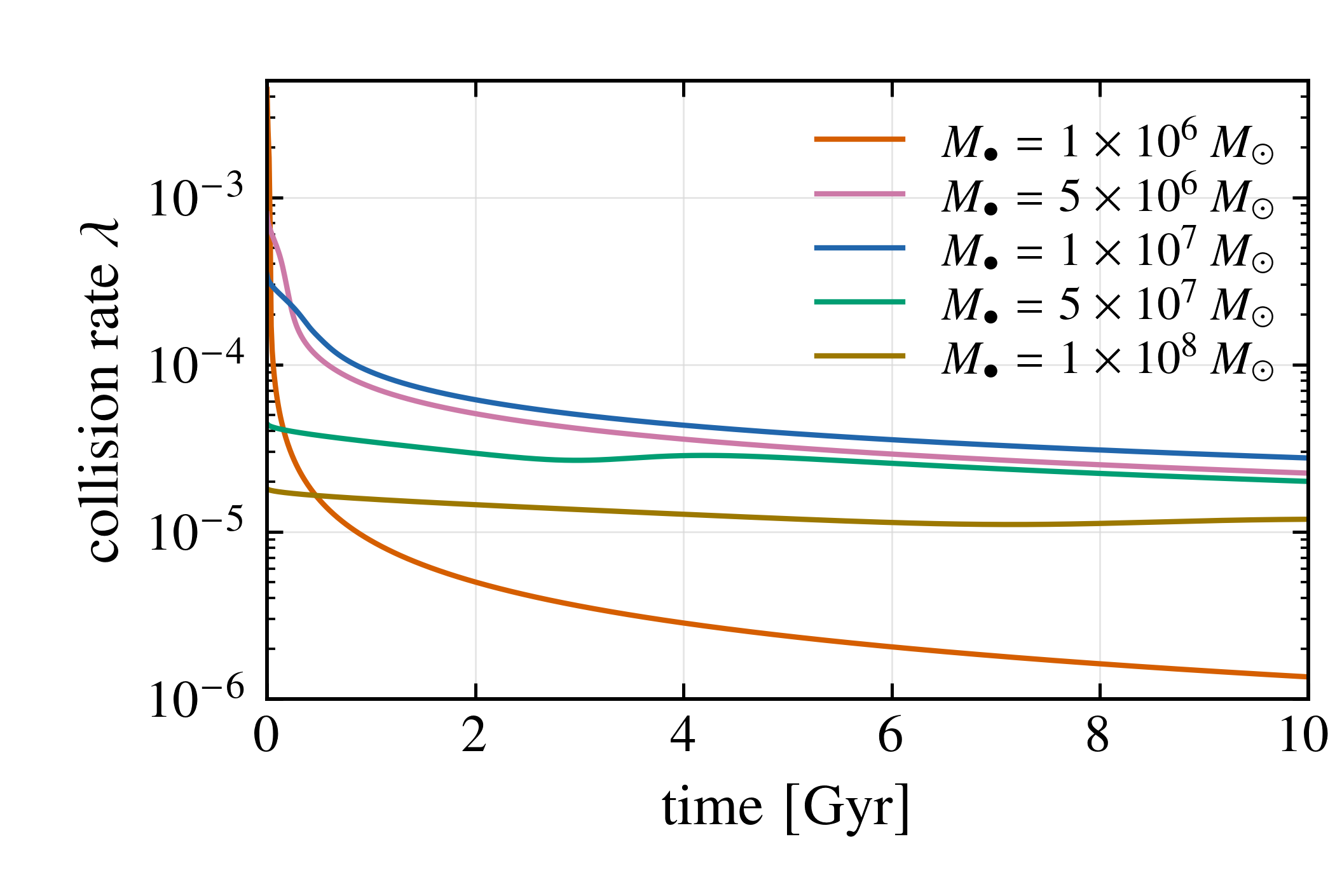}
    \caption{Stellar collision rates experienced by a single target star on a circular orbit at $r_0=100M_\bullet$ and for different host supermassive black holes. The evolution of the source frame collision rate $\lambda$ (in ${\rm yr}^{-1}$) is presented as a function of time. Lower-mass models initially have higher rates but undergo earlier and stronger declines because of the depletion of stars. }
    \label{fig:collision_rate}
\end{figure}

Deep in the gravitational potential of the black hole, the relative velocities of colliding stars can reach $\sim 0.1c$, giving rise to energetic explosions~\cite{HuLoeb2024Energetic}.
Using radiation hydrodynamics simulations, we investigate the resulting electromagnetic signatures, including hour-long X-ray bursts powered directly by the collisions, possible optical and radio counterparts, and additional emission from subsequent tidal disruption and accretion of the collision debris. These transients may serve as independent probes of stellar populations and gas environments near supermassive black holes.

\vspace{0.2cm}

\noindent{ \it Population analysis -}
Previous Fokker–Planck and Monte Carlo studies have investigated stellar collisions in nuclear star clusters surrounding massive black holes~\cite{MurphyCohnDurisen1991,FreitagBenz2002Nuclei}.
The rate of destructive encounters depends sensitively on the supply of stars to tightly bound orbits~\cite{BalbergYassur2023}. 
Here we calculate the starclash rate experienced by a single target star on a circular orbit of radius $r_0$.



 Our cluster model consists of 16 stellar mass components drawn from a Kroupa initial mass function over $0.08$--$100 \ M_\odot$~\cite{Kroupa2001}. The different stellar components contribute to gravitational scattering and mass segregation, modifying their distribution in the inner cluster~\cite{AlexanderHopman2009, Balberg2024Segregation}. All components initially follow the same Tremaine density profile~\cite{Tremaine1994}, with a total stellar mass of $20M_\bullet$ (see Supplemental Material, hereafter SM). This model has previously been applied to calculate the time-dependent rates of extreme mass-ratio inspirals in nuclear star clusters~\cite{Sun2026EMRI, PanLyuYang2021}.

The orbital distribution of each component is described by $f_i(E,R,t)$, where $i$ denotes different stellar components, $E=\phi(r)-v^2/2$ is the positive specific binding energy, and $R=J^2/J_c^2(E)$ is the normalized squared angular momentum. The orbit-averaged Fokker–Planck equation takes the flux-conservative form~\cite{CohnKulsrud1978,PanYang2021}
\begin{align}
    C(E,R) \frac{\partial f_i}{\partial t} = -\frac{\partial F_{E,i}}{\partial E} - \frac{\partial F_{R,i}}{\partial R}
\end{align}
where $F_{E,i}$ and $F_{R,i}$ are the energy and angular momentum fluxes, and $C(E,R)=4\pi^2P(E,R)J_c^2(E)$ is the phase-space volume factor, with $P(E,R)$ being the radial orbital period.  The mass components are coupled through the diffusion coefficients describing gravitational encounters. The flux expressions and the initial and boundary conditions are given in the SM.

\begin{figure*}
    \centering
    \includegraphics[width=0.8\textwidth]{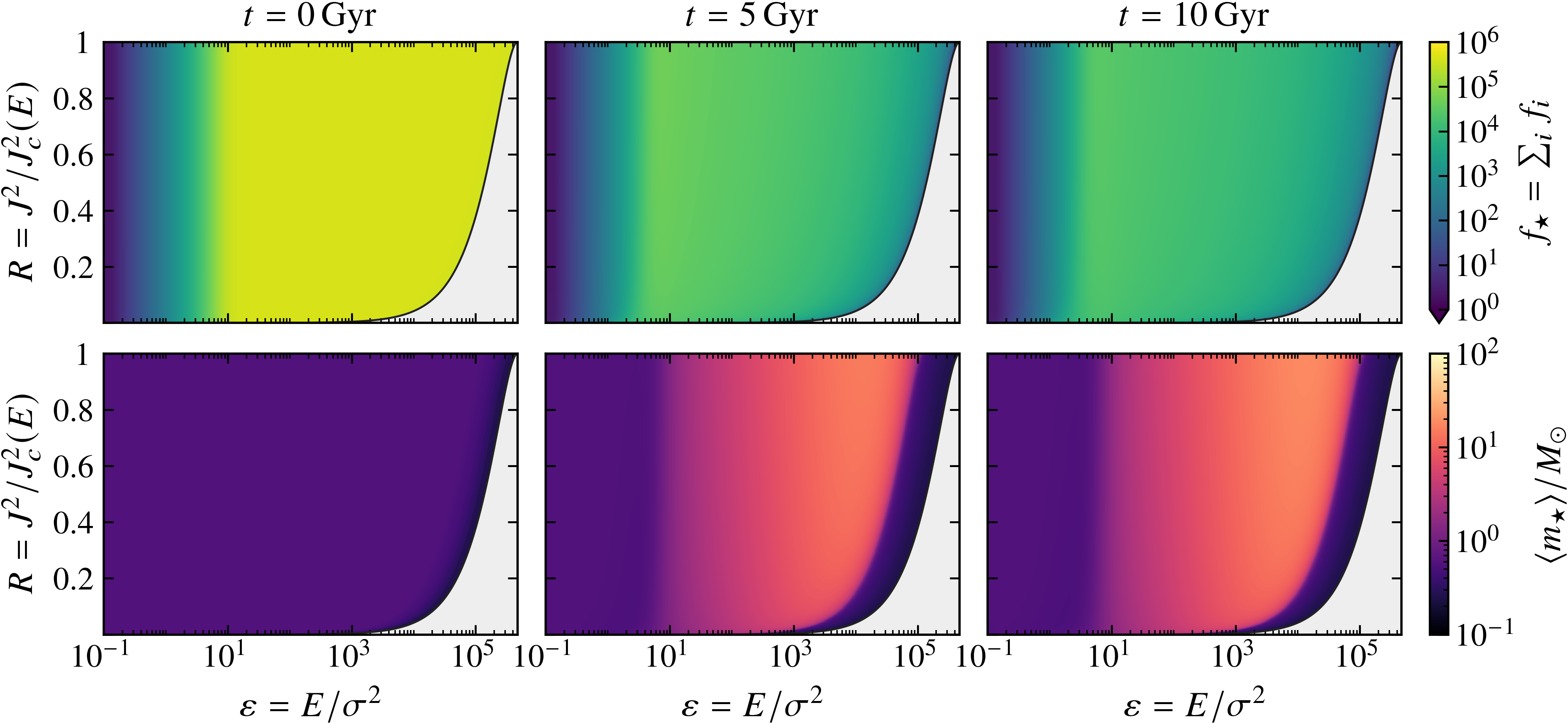}
    \caption{Evolution of the stellar phase-space distribution in a nuclear star cluster surrounding a black hole of mass \(M_\bullet=10^7\,M_\odot\). From left to right, the columns show \(t=0\), \(5\), and \(10\,\mathrm{Gyr}\). The axes are the dimensionless binding energy \(\varepsilon=E/\sigma^2\) and the normalized angular momentum \(R=J^2/J_c^2(E)\). The upper panels show the total stellar number distribution function \(f_\star=\sum_{i=1}^{16}f_i\), summed over the 16 stellar mass components. The lower panels show the mean stellar mass at each point in phase space, \(\langle m_\star\rangle=\sum_i m_i f_i/f_\star\). Gray regions lie outside the modeled stellar distribution. As the cluster evolves, \(f_\star\) decreases at high binding energies while the mean stellar mass there increases, reaching local maxima of \(15\,M_\odot\) at \(5\,\mathrm{Gyr}\) and \(18\,M_\odot\) at \(10\,\mathrm{Gyr}\). This concentration of massive stars on tightly bound orbits is the signature of mass segregation.}
    \label{fig:star_distribution}
\end{figure*}

We evolve nuclear star clusters surrounding massive black holes of $M_\bullet=10^6-10^8M_\odot$ over 10 Gyr. Fig.~\ref{fig:star_distribution} shows the evolution of the stellar distribution for $M_\bullet=10^7M_\odot$. At high binding energies, the total stellar distribution declines while the local mean stellar mass increases as massive stars segregate onto tightly bound orbits.

A cluster star on an orbit characterized by $(E,R)$ can encounter the target only if its orbit crosses $r_0$. Its tangential and radial velocities at $r_0$ satisfy
\begin{align}
    v_\perp^2(E,R;r_0) &= \frac{R J_c^2(E)}{r_0^2} \notag \\
    v_r^2(E,R;r_0) &= 2[\phi(r_0)-E]-v_\perp^2(E,R;r_0)
\end{align}


During one radial period, the star spends a total time $2\mathrm{d}r/|v_r|$ within a spherical shell of width $\mathrm{d}r$ at $r_0$. The contribution of the corresponding orbital population to the local stellar number density is therefore
\begin{align}
    \mathrm{d}n_{\rm star} = \frac{2C(E,R)f_{\rm star}(E,R,t)}{4\pi r_0^2 P(E,R) |v_r|} \mathrm{d}E\,\mathrm{d}R
\end{align}

Let $\varphi$ denote the angle between the tangential velocity of the incident star and the velocity of the circular orbit target. Their relative speed is 
\begin{align}
    w^2 = v_r^2 + v_\perp^2 + v_c^2 - 2v_\perp v_c \cos\varphi
\end{align}
For a spherical, nonrotating stellar distribution, we average over the uniformly distributed angle $\varphi$. Gravitational focusing is negligible at the encounter velocities considered here~\cite{RubinLoeb2011}, and we adopt the geometric collision cross section $\Sigma_{\rm coll}=\pi(R_{\star,1}+R_{\star,2})^2$. Integrating $\mathrm{d}\lambda=\mathrm{d}n_{\rm star}\Sigma_{\rm coll}\langle w\rangle_\varphi$ over the allowed stellar phase space yields
\begin{equation}
    \lambda(r_0,t;M_\bullet) = \frac{2\Sigma_{\rm coll}}{4\pi r_0^2} \int\limits_{\mathcal{D}(r_0)}  \mathrm{d}E\,\mathrm{d}R \frac{C f_{\rm star}(t)}{P} 
    \frac{\langle w( r_0) \rangle_\varphi}{|v_r(r_0)|} 
\label{eq:single_target_collision_rate}
\end{equation}
where $\mathcal D(r_0)={(E,R):r_{\rm tidal}<r_p(E,R)\leq r_0\leq r_{\rm apo}(E,R)}$, $r_p$ and $r_{\rm apo}$ are the orbital pericenter and apocenter respectively and $r_{\rm tidal}$ is the tidal disruption radius of the incident star.

Fig.~\ref{fig:collision_rate} shows the time-dependent collision rates for a target at $r_0=100GM_\bullet/c^2$. Lower-mass black holes initially exhibit higher collision rates but also earlier and stronger declines due to the depletion of stars on tightly bound orbits. The rates evolve more slowly in more massive systems, consistent with their longer relaxation times~\cite{Broggi2022Rates}. A similar decline in the cluster-wide collision rate has been found in simulations following stellar encounters around growing black holes~\cite{ZhangAmaroSeoane2026Collisions}.

Adopting the massive black hole mass functions $\Phi(M_\bullet)=\mathrm{d}n/\mathrm{d}\log M_\bullet$ from~\cite{Babak2017} and assuming that $10\%$ of massive black holes host one such close companion, we estimate cumulative event rates of $\sim 2-3\times10^3 \ {\rm yr}^{-1}$ out to $z=1$ for the two benchmark mass functions (see SM). Within $z=0.1$, the estimated rate  is order of  $10{\rm yr}^{-1}$.

\begin{figure*}[!tbp]
\centering
\includegraphics[width=0.8\textwidth]{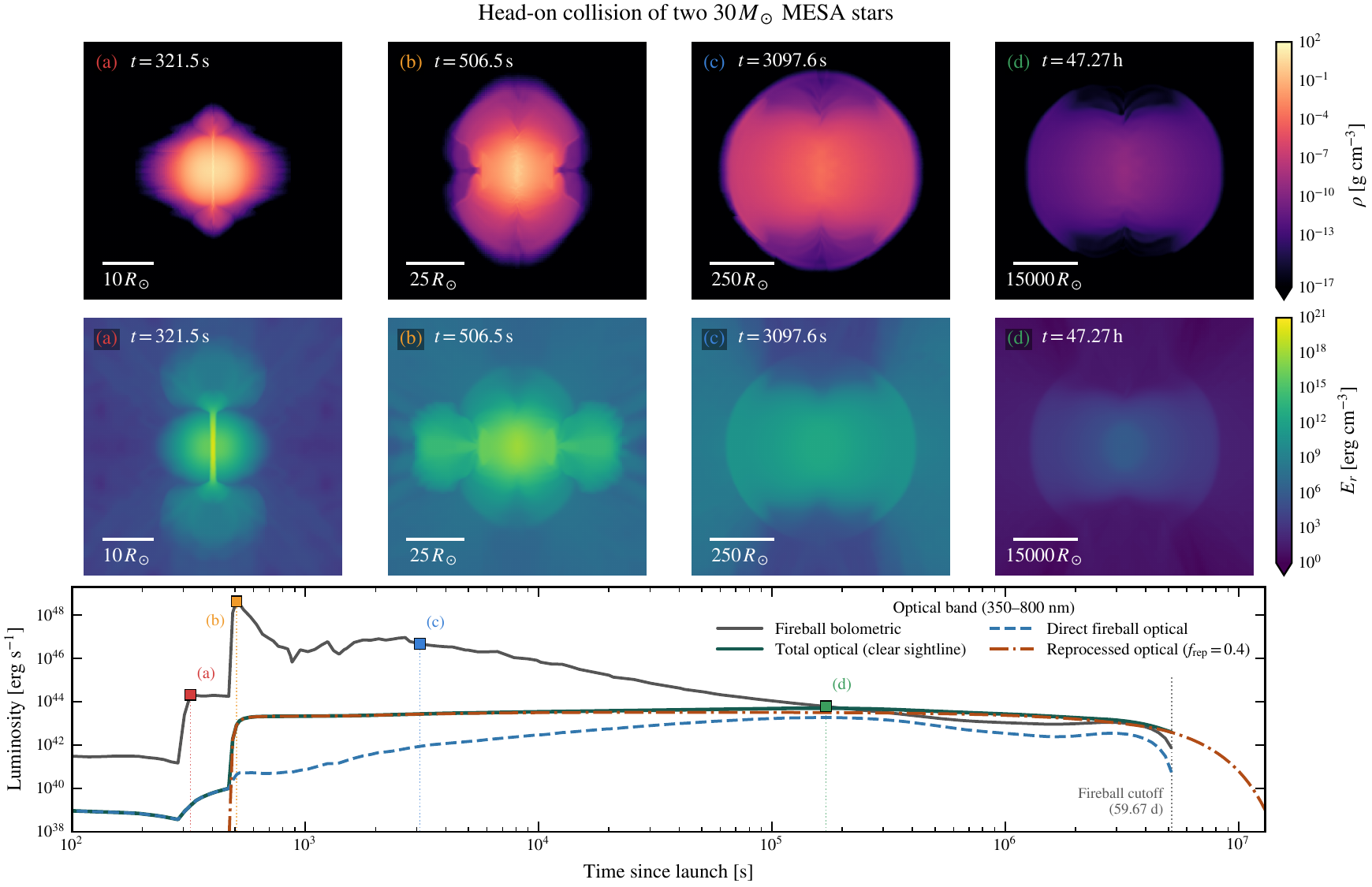}
\caption{Fiducial head-on collision. The first row shows density and the
second row lab-frame radiation energy density in the collision plane
$z=0$, at the four snapshots labeled inside each panel. White
segments show lengths of $10$, $25$, $250$, and $15000R_\odot$ from left
to right in both rows. 
Panels (a)–(c) illustrate three successive stages of shock evolution, corresponding to three enhancements in the X-ray luminosity, while panel (d) shows the late-time expansion of the ejecta.
The bottom
row shows the bolometric $L_{\bar\tau=1}$ light curve; colored squares
and matching labels identify exactly the same four snapshots.}
\label{fig:nu_fiducial_evolution}
\end{figure*}

\vspace{0.2cm}
\noindent{\it Collision simulation -} Previous hydrodynamic studies have investigated mass loss and disruption in high-velocity stellar collisions~\cite{FreitagBenz2005Collisions}, while radiation-transfer calculations have predicted UV and optical light curves for collisions involving red giants~\cite{Ryu2024Giants,Dessart2024Giants} and main-sequence stars~\cite{RyuDessart2025MS}. To resolve the prompt emission from collisions at higher velocities, we perform three-dimensional radiation hydrodynamics (RHD) simulations of isolated stellar encounters. We consider two sets of simulations to explore the dependence on encounter geometry and stellar structure. The first consists of collisions between identical solar-mass, solar-radius polytropes with normalized impact parameters $\tilde b=0,0.1,0.3,0.5,$ and $0.7$. The second consists of head-on collisions between identical zero-age main-sequence stars constructed with MESA~\cite{Paxton2011,Jermyn2023}, with individual masses of $1,5,10,20,$ and $30M_\odot$. In all simulations, each star is launched at $0.1c$ in the center-of-mass frame, corresponding to a relative velocity of $0.2c$ in our Newtonian treatment.


We use AthenaK~\citep{Stone2026} with the self-gravity, stellar initialization, and remapping infrastructure described in~\citep{2026arXiv260937859J}. The Newtonian hydrodynamic equations are coupled to grey radiation moment equations:
\begin{align}
 \partial_t\rho+\nabla\cdot(\rho\boldsymbol v)&=0,\nonumber\\
 \partial_t(\rho\boldsymbol v)
 +\nabla\cdot(\rho\boldsymbol v\boldsymbol v+p_g\mathsf I)
 &=\boldsymbol f_g+\boldsymbol G,\nonumber\\
 \partial_t E_g+\nabla\cdot[(E_g+p_g)\boldsymbol v]
 &=\boldsymbol v\cdot\boldsymbol f_g+cG^0-\mathcal L_\nu,\nonumber\\
 \partial_t \mathsf{E}_r+\nabla\cdot\mathsf{F}_r&=-cG^0,\nonumber\\
 c^{-2}\partial_t\mathsf{F}_r+\nabla\cdot\mathsf{P}_r&=-\boldsymbol G.
 \label{eq:nu_rhd}
\end{align}
Here $E_g=u_g+\rho v^2/2$ is the gas energy density, $\boldsymbol{f}_g$ is the gravitaional acceleration and $\mathsf{E}_r$, $\mathsf{F}_r$, and $\mathsf{P}_r$ are the radiation energy density, flux, and pressure tensor in the lab frame. The gas pressure $p_g$ follows a tabulated H/He equation of state~\cite{Tomida2013,Chabrier2021,Timmes2000}. The radiation four-force $(cG^0,\boldsymbol G)$ describes matter–radiation coupling, while $L_\nu$ accounts for thermal neutrino losses~\cite{Itoh1996}. The radiation closure blends M1~\citep{Levermore1984} with angular information obtained from an auxiliary stationary transfer calculation. Further numerical details are given in the SM.

Fig.~\ref{fig:nu_fiducial_evolution} illustrates the evolution of a head-on collision between two \(30M_\odot\) stars. The characteristic collision timescale, \(t_{\rm col}\sim 2R_\star/v_{\rm rel}\), is much shorter than the subsequent radiative evolution. The collision drives strong shocks that convert bulk kinetic energy into internal and radiation energy. To estimate the band-limited luminosities, we weight the radiative flux on the angle-averaged photosphere by the corresponding blackbody spectral fraction at the local radiation temperature (see SM).

The early shock evolution proceeds through three distinct stages, each associated with an enhancement in the X-ray luminosity. First, the collision-driven shocks break out perpendicular to the collision plane (Panel a). The shocked gas subsequently expands and collides near the equatorial plane, driving a secondary shock that breaks out along the plane (Panel b). As this shock propagates outward and overtakes the previously expanding gas, renewed collisions lead to further shock heating and a third enhancement in the X-ray luminosity (Panel c). The ejecta eventually approach adiabatic expansion (Panel d), with radiation remaining trapped in regions where the photon diffusion time exceeds the expansion time~\cite{arnett1980analytic} and gradually escapes as the ejecta expand and become optically thin.

\begin{figure}[!tbp]
\includegraphics[width=0.95\linewidth]{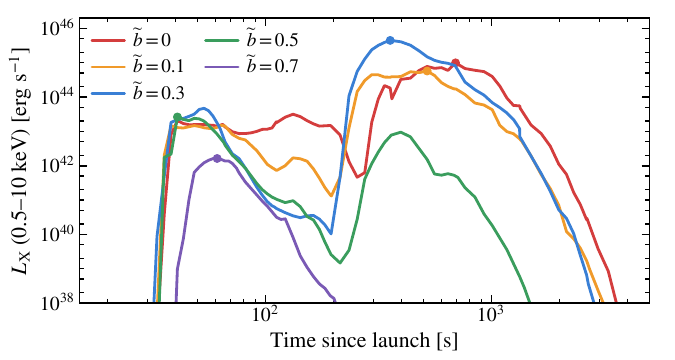}
\includegraphics[width=0.95\linewidth]{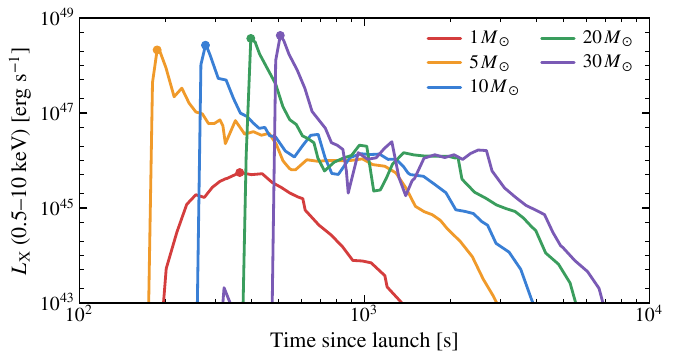}
\caption{X-ray luminosities of stellar collisions with different impact parameters $\widetilde{b}$ (top panel) and stellar masses (bottom panel). Dots mark individual luminosity peaks. The impact-parameter sequence uses pairs of polytropes, each with a mass of $1 M_\odot$ and a radius of $1 R_\odot$. The stellar-mass sequence consists of head-on collisions between pairs of ZAMS stellar models constructed with MESA at the sampled initial masses.}
\label{fig:Xray_light_curves}
\end{figure}

The high collision velocity leads to substantial shock heating: the initial kinetic energy per baryon is $m_pv^2/2\simeq4.7,{\rm MeV}$ for $v=0.1c$. The escaping radiation, however, is regulated by successive shock breakouts, photon diffusion, and adiabatic cooling. Assuming local blackbody emission at the photosphere, we reconstruct the emission spectra and calculate the X-ray luminosities in the 0.5–10 keV band, as shown in Fig.~\ref{fig:Xray_light_curves}. The massive-star models exhibit a rapid rise in X-ray luminosity on timescales of $\mathcal{O}(10^2-10^3)$~s, associated with the early shock breakouts~\cite{nakar2010breakout}. 
The absence of a comparable early rise in the $1M_\odot$ model may be attributed to a larger overlying mass column, which delays radiation escape and allows more adiabatic cooling. The peak X-ray luminosity increases substantially with stellar mass, spanning approximately $10^{45}-10^{48}$~ erg/s across the stellar-mass sequence, with the bursts lasting up to several hours.

Interestingly, the peak X-ray luminosity does not decrease monotonically with increasing impact parameter. For the polytropic models, the collision with $\tilde b=0.3$ produces a peak luminosity approximately $4.5$ times that of the head-on collision, possibly because radiation escapes more efficiently along directions of lower column density. More grazing encounters produce substantially weaker X-ray emission owing to the smaller overlap between the stars and weaker shock heating.

As the fireball expands and cools, its emission gradually shifts toward lower frequencies. For the head-on collision between two $30M_\odot$ stars, the bolometric luminosity peaks at approximately $500$~s, with the emission primarily in the X-ray band (Fig.~\ref{fig:nu_fiducial_evolution}). The subsequent optical emission from the fireball for different collision configurations is discussed in the SM.

\vspace{0.2cm}
\noindent{\it Secondary emission -} Stellar mass loss from the nuclear star cluster can supply clumpy gas inflows and a reservoir for later AGN activity~\cite{Schartmann2010Fuel}. Such gas may intercept and reprocess part of the radiation from the collision-induced fireball. For a gas cloud of characteristic size $R_{\rm rep}$ and electron density $n_e$ near a $10^7 M_\odot$ massive black hole, the photon diffusion timescale is approximately
\begin{align}
t_{\rm esc} &\sim \frac{\sigma_T n_e R^2_{\rm rep}}{c}\nonumber\\
&\sim 17 {\rm days} \left ( \frac{n_e}{3\times 10^{10}{\rm cm^{-3}}} \right ) \left (\frac{R_{\rm rep}}{10^3 GM_\bullet/c^2} \right )^2
\end{align}
where $\sigma_T$ is the Thomson cross section. 
Following effective models of radiative reprocessing in optically thick gas~\cite{roth2016reprocessing,piro2020wind},we describe the energy evolution of the reprocessing medium by
\begin{align}
\frac{d E}{d t} = f_{\rm rep} L_{\rm fb} -\frac{E}{t_{\rm esc}}
\end{align}
where $L_{\rm fb}$ is the bolometric luminosity of the fireball, $f_{\rm rep}$ is the fraction of radiation absorbed by the surrounding gas.
Assuming thermal emission, the effective temperature is estimated from $4\pi R_{\rm rep}^2\sigma_{\rm SB}T_{\rm eff}^4=E/t_{\rm esc}$.
Fig.~\ref{fig:nu_fiducial_evolution} illustrates the resulting optical light curve for $f_{\rm rep}=0.4$.
The reprocessed emission may dominate over the direct optical emission from the fireball at late times, sustaining an optical transient over several diffusion timescales.

The interaction of the expanding fireball with the circumnuclear gas may also produce synchrotron radio emission, analogous to the radio counterparts of tidal disruption events \citep{Yalinewich:2019oyh,Matsumoto:2021gcg,Zhuang:2024tos}. The radio emission may emerge at later times as the ejecta expand and free–free absorption by the surrounding gas becomes less significant.

Depending on its post-collision trajectory, part of the expanding debris may undergo subsequent tidal disruption by the massive black hole. Fallback, stream collisions, and accretion of the bound debris may power additional emission ~\cite{Balberg2013Collisions,hu2024collision,Brutman2024Streams}, potentially giving rise to a delayed UV or soft X-ray flare.


\vspace{0.2cm}
\noindent{\it Discussion -} In our model, shocks from a stellar collision power an hour-long starclash X-ray burst independently of accretion onto the supermassive black hole. This model assumes that a fraction of supermassive black holes host close orbiting stars, as motivated by the star--disk collision interpretation of QPEs~\cite{LinialMetzger2023QPE,Zhou2024QPEPopulation}. Extended emission-line regions in QPE hosts also support a past AGN phase in which such stars could have migrated inward~\cite{Wevers2024Hosts,Xiong2025,JiangPan2025Embers,Wevers2026}. These stars have a significant chance to collide with a cluster star in the timescale of $10^4-10^5$ years. The collisions are expected to be relativistic because the stars are deep in the gravitational potential of the supermassive black hole. Therefore the kinetic energy per baryon may exceed MeV and the post-collision fireball is capable of producing a strong X-ray burst. Because of the special environment near the supermassive black hole, we may expect reprocessed emission from the environmental gas clouds that later dominates in the optical band or even another tidal disruption event.

The population study reveals that the event rate within redshift one may reach $[2-3]\times 10^3$ per year. Therefore it is a promising target to search for wide-field X-ray telescopes (such as the Einstein Probe~\cite{Yuan2025EinsteinProbe}) and perform optical followups if candidate starclash X-ray bursts from galactic centers are found. Assuming a population of such events observed, the inferred rate and energetics of high-velocity stellar collisions could constrain the density and orbital distribution of stars near supermassive black holes, building up the synergy with observations of quasi-periodic eruptions and future gravitational-wave detections of extreme mass-ratio inspirals~\cite{Babak2017}: "QPE-EMRI-SXB". Where surrounding gas reprocesses the collision radiation, the delay and spectrum of the optical and ultraviolet emission could also constrain the gas geometry and column density in otherwise quiescent galactic nuclei.


\begin{acknowledgments}
We thank Zhen Pan for discussions about quasi-periodic eruptions.
HY is supported by the Natural Science Foundation
of China (Grant 12573048).
\end{acknowledgments}  

\appendix

\section{Fokker-Planck Modeling of the Population}
\label{app:population}

\subsection{Initial cluster and stellar mass function}

We model a spherical, nonrotating nuclear star cluster around a massive
black hole (MBH) of mass $M_\bullet$. The initial stellar mass is
$M_{\rm cl}=20M_\bullet$. All stellar components follow a Tremaine
profile~\cite{Tremaine1994} with inner slope $\gamma=3/2$ and scale
radius $r_a=4GM_\bullet/\sigma^2$. We set the velocity dispersion $\sigma$
by the adopted $M_\bullet$--$\sigma$ relation~\cite{Gultekin2009,PanYang2021},
\begin{equation}
 M_\bullet=1.53\times10^6M_\odot
 \left(\frac{\sigma}{70\,\mathrm{km\,s^{-1}}}\right)^{4.24}.
\end{equation}
In the Tremaine's model, the stars number density and cluster total potential are
\begin{align}
 n(r,t=0)&=\frac{(3-\gamma)N}{4\pi}
        \frac{r_a}{r^\gamma(r+r_a)^{4-\gamma}},\\
 \phi(r)&=\frac{GM_\bullet}{r}
 +\frac{GM_{\rm cl}}{(2-\gamma)r_a}
 \left[1-\left(\frac{r}{r+r_a}\right)^{2-\gamma}\right].
 \label{eq:sm_potential}
\end{align}
where $N$ is the stars number.

Defining $\mu=m/M_\odot$, we adopt a Kroupa initial mass
function (IMF)~\cite{Kroupa2001},
\begin{equation}
 {\cal I}(\mu)\equiv\frac{dN}{d\mu}=A
 \begin{cases}
   \mu^{-1.3},      &0.08\leq\mu<0.5,\\
   0.5\,\mu^{-2.3}, &0.5\leq\mu\leq100 .
 \end{cases}
 \label{eq:sm_imf}
\end{equation}
The normalization satisfies
$\int_{0.08}^{100}\mu{\cal I}(\mu)\,d\mu=M_{\rm cl}/M_\odot$. We divide the total cluster into $16$ components, and all components mass edge, in solar mass, are
\begin{equation*}
 \begin{gathered}
 0.08,\ 0.12,\ 0.20,\ 0.35,\ 0.50,\ 0.80,\ 1.3,\ 2,\ 3.5,\\
 5,\ 8,\ 12,\ 20,\ 30,\ 50,\ 75,\ 100 .
 \end{gathered}
\end{equation*}
For each component $i$, we use a representative mass to describe this component, the mass and number satisfy
\begin{align}
    N_i &= \int_{\mu_i}^{\mu_{i+1}}{\cal I}(\mu)\,d\mu, \notag \\
    m_i &= \frac{1}{N_i} \int_{\mu_i}^{\mu_{i+1}} \mu \mathcal{I}(\mu) \ d\mu  \ M_\odot
\end{align}

This prescription preserves the number and mass in each bin. The stellar radius satisfy
\begin{equation}
 \frac{R_{\star,i}}{R_\odot}=
 \begin{cases}
    \mu^{0.8}, &\mu\leq 1,\\
    \mu^{0.57},&\mu>1 
 \end{cases}
\end{equation}

We treat the IMF as the initial condition for a population in which
stellar births balance finite-lifetime deaths within each mass bin. During the cluster evolution, we don't consider the stellar birth and death.

\subsection{Orbital-averaged transport and boundary conditions}

We use orbital averaged Fokker Planck (FP) equation to consider the cluster evolution in the $(E,R)$ phase space. $E=\phi(r)-v^2/2$ is the positive specific energy and $R=J^2/J_c^2(E)$ is the normalized squared angular momentum, where $J_c(E)$ is the specific angular momentum of a circular orbit at energy $E$.

The FP equation can be expressed 
\begin{equation}
    C\frac{\partial f_i}{\partial t} = - \frac{\partial F_{E,i}}{\partial E} - \frac{\partial F_{R,i}}{\partial R}
\end{equation}
where $F_{E,i}$ and $F_{R,i}$ are the fluxes of each component along the energy and angular momentum direction, respectively.

The number density in phase space is $dN_i = C f_i dEdR$, with $C=4\pi^2 P(E,R) J_c^2(E)$, where $P(E,R)$ is the orbital radial period. The initial isotropic distribution function $f_i(E)$ can be obtained from the initial number density $n_i(r)$ using Eddington inversion
\begin{equation}
    f_{i}(E,t=0) = \frac{1}{\sqrt{8} \pi^2} \frac{d}{dE} \int_0^E \frac{dn_i}{d\phi} \frac{d\phi}{\sqrt{E-\phi}}
\end{equation}

The fluxes can be written into
\begin{align}
    -F_{E,i} &= D_{EE,i} \frac{\partial f_i}{\partial E} + D_{ER,i} \frac{\partial f_i}{\partial R} + D_{E,i} f_i \notag \\
    -F_{R,i} &= D_{RR,i} \frac{\partial f_i}{\partial R} + D_{ER,i} \frac{\partial f_i}{\partial E} + D_{R,i} f_i
\end{align}
where $\{D_{EE,i}, D_{RR,i}, D_{ER,i}\}$ are the diffusion coefficients and $\{D_{E,i}, D_{R,i}\}$ are the advection coefficients for each component due to the gravitational scattering. The precise expreesion of these coefficients can be obtained in ~\cite{PanYang2021}.

In our Fokker Planck simulations, the normalized angular momentum range is $R_{{\rm lc},i} \leq R \leq 1$. $R_{{\rm lc},i}$ is the so called loss cone boundary. For each star component, the loss cone boundary satisfy where the orbital pericenter is
\begin{align}
    r_{\rm lc,i} = \max \left[R_{\star,i} \left(\frac{M_\bullet}{m_i}\right)^{1/3}, 8r_g\right]
\end{align}
where $r_g$ is the MBH gravitational radius. The first term is the tidal disruption radius and the second term is to match the last stable orbit in general relativity cases.

The loss cone boundary is
\begin{align}
    R_{\rm lc,i}(E) = \frac{2 r_{\rm lc,i}^2 [\phi(r_{\rm lc,i})-E]}{J_c(E)^2}
\end{align}

The energy range in our simulations is $0.1\sigma^2 \leq E \leq 0.999E_{t,i}$, and the place $E_{t,i}$ satisfy $R_{{\rm lc},i}(E_{t,i})=1$. 

For the boundary condition, at minimum energy boundary we use $f_i(E=0,t)=f_i(E=0,t=0)$, at maximum energy boundary we choose $F_{E,i}=0$, and at maximum angular momenmtum boundary we choose $F_{R,i}=0$. For loss cone boundary, we use Cohn-Kulsrud condition~\cite{CohnKulsrud1978, PanLyuYang2021}. Defining $\nu_i=D_{RR,i}/(C R_{{\rm lc},i})$ and $q_i=P\nu_i/R_{{\rm lc},i}$, the Cohn-Kulsrud condition can be written into
\begin{align}
    R'_{{\rm lc},i}(E) F_{E,i} - F_{R,i} &= \frac{C \nu_i}{B(q_i)} f_i(E,R_{{\rm lc},i},t) \notag \\
    B(q) &= 
    \begin{cases}
        0.824 \sqrt{q}, &q<1, \\
        q, & q\geq 1
    \end{cases}
\end{align}

\subsection{Collision rate for a circular target}

We place a target of mass $m_{\rm tar} = 1M_\odot$ and radius $R_{\rm tar}=1R_\odot$ on
a circular orbit at $r_0=100r_g$. A field star in component $i$
contributes if its orbit belongs to its modeled domain and satisfies
$r_{{\rm loss},i}<r_p\leq r_0\leq r_{\rm apo}$. We denote this set by
$\mathcal D_i(r_0)$. A component with $r_{{\rm loss},i}\geq r_0$ gives
zero rate. At $r_0$ the field star has tangential speed
$v_\perp=J_c(E)\sqrt{R}/r_0$. In one radial period it spends
$2\,dr/|v_r|$ in a shell of width $dr$, counting both passages.
By spherical symmetry the expected number in the shell divided by its
volume gives the local density at the target,
\begin{equation}
 dn_i=\frac{C f_i\,dE\,dR}{4\pi r_0^2 dr}
       \frac{2\,dr}{P|v_r|}
     =\frac{2\pi J_c^2(E) f_i}{r_0^2|v_r|}\,dE\,dR .
\end{equation}
Let $\varphi$ denote the angle between the field star's tangential
velocity and the target velocity in the local tangent plane. This angle
is uniform in a spherical, nonrotating cluster. With $v_c=v_c(r_0)$,
\begin{align}
 w^2(\varphi)&=v_r^2+v_\perp^2+v_c^2-2v_\perp v_c\cos\varphi,\\
 \langle w\rangle_\varphi&=\frac{1}{2\pi}\int_0^{2\pi}w(\varphi)\,d\varphi .
\end{align}
The geometric contact cross section is
$\Sigma_i=\pi(R_{\rm tar}+R_{\star,i})^2$. Integrating
$dn_i\,\Sigma_i\langle w\rangle_\varphi$ gives the source-frame rate
per target at $(r_0,t;M_\bullet)$,
\begin{align}
 \lambda&=\sum_{i=1}^{16}\lambda_i, \notag \\
 \lambda_i&=\frac{2\pi\Sigma_i}{r_0^2}
 \int_{\mathcal D_i(r_0)}
 J_c^2(E) f_i(E,R,t)\,
 \frac{\langle w\rangle_\varphi}{|v_r|}\,dE\,dR .
 \label{eq:sm_collision}
\end{align}

\begin{figure}[t]
 \centering
 \includegraphics[width=\linewidth]{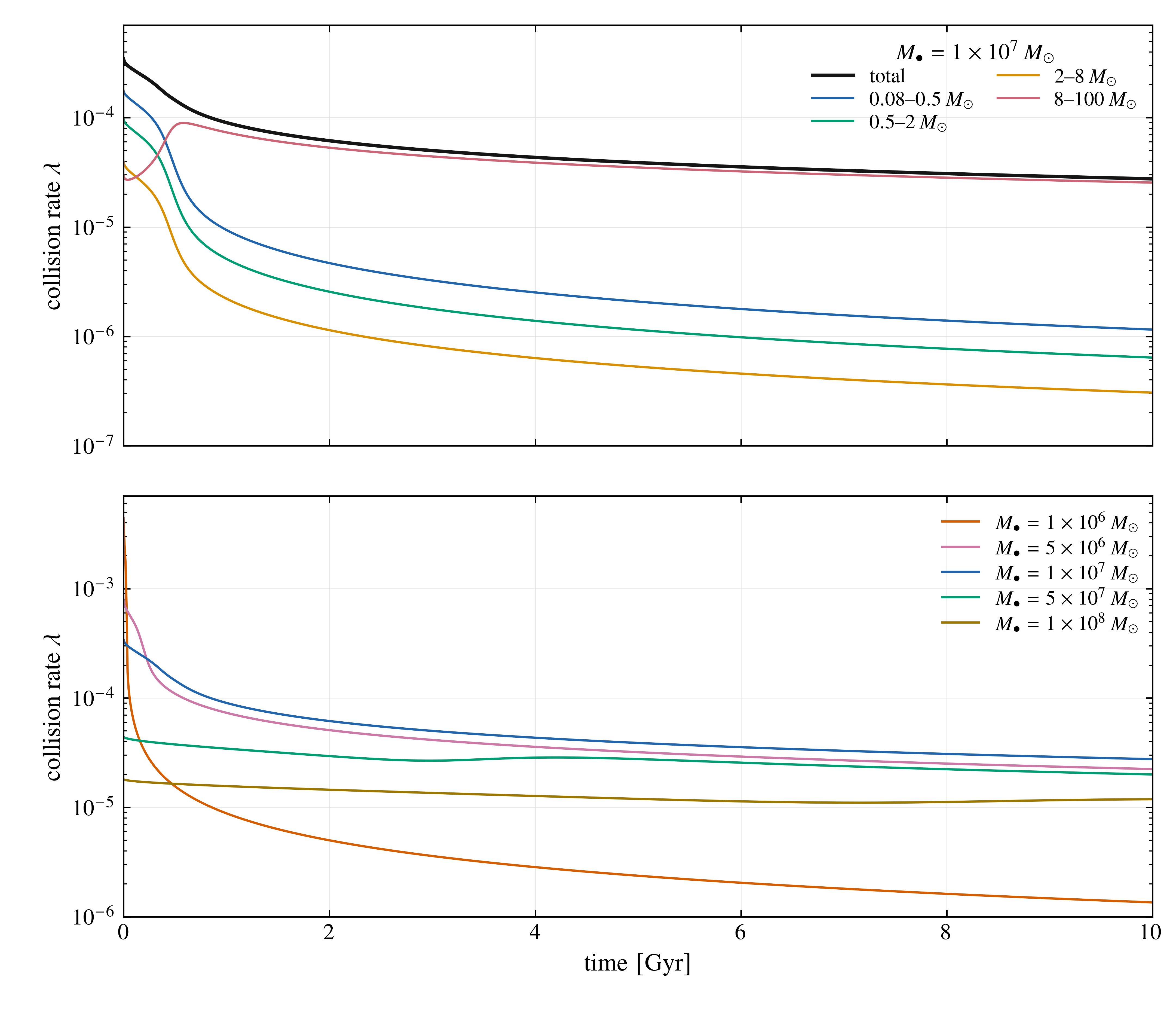}
 \caption{Source-frame geometric collision rates for one solar-type
 target at $100r_g$. The upper panel shows the total rate for
 $M_\bullet=10^7M_\odot$ and the contributions from four stellar mass
 intervals, obtained by summing the 16 simulated components. The lower
 panel shows total rate histories for five representative MBH masses.}
 \label{fig:sm_rate_components}
\end{figure}

Figure~\ref{fig:sm_rate_components} shows the role of the stellar mass
distribution. For $M_\bullet=10^7M_\odot$, stars below $0.5M_\odot$
supply $51.3\%$ of the initial rate. Stars in the $8$--$100M_\odot$
interval represent $0.63\%$ of the initial population by number and
supply $92.4\%$ of the rate at $10\,\mathrm{Gyr}$. This is the reason of mass segregation.

\subsection{Cosmological rate}

For each MBH mass we use the time-averaged rate
\begin{equation}
 \bar\lambda(M_\bullet)=\frac{1}{T}\int_0^T
          \lambda(100r_g,t;M_\bullet)\,dt,\qquad T=10\,\mathrm{Gyr}.
\end{equation}
Table~\ref{tab:sm_geometric_means} shows the averaged rates in our different  clusters simulations. 
We adopt the same age-averaged rate at each redshift to consider the star collision cosmological rate.
\begin{table}[t]
 \caption{Mean source-frame geometric collision rate for one solar-type
 target on a circular orbit at $100r_g$, averaged over $0$--$10\,\mathrm{Gyr}$.}
 \label{tab:sm_geometric_means}
 \centering
 \begin{tabular}{rr}
 \hline\hline
 $M_\bullet/(10^6 M_\odot)$ & $\bar\lambda/(10^{-5}\,\mathrm{yr}^{-1})$\\
 \hline
 1   & 1.198\\
 3   & 3.657\\
 5   & 4.992\\
 8   & 5.467\\
 10  & 5.447\\
 30  & 3.830\\
 50  & 2.702\\
 80  & 1.680\\
 100 & 1.282\\
 \hline\hline
 \end{tabular}
\end{table}

We adopt the two comoving MBH mass
functions of Ref.~\cite{Babak2017} and extend the power laws over
$10^6M_\odot\leq M_\bullet\leq10^8M_\odot$,
\begin{equation}
 \Phi_\pm(M_\bullet)\equiv\frac{dn_{\rm com}}{d\log(M_\bullet/M_\odot)}
     = A_\pm\left(\frac{M_\bullet}{3\times10^6M_\odot}\right)^{\alpha_\pm}.
\end{equation}
The pairs $(\alpha_-,A_-)$ and $(\alpha_+,A_+)$ are $(-0.3,0.005)$ and
$(+0.3,0.002)$, respectively, with $A_\pm$ in
$\mathrm{Mpc}^{-3}$. We hold both mass functions fixed
over $0\leq z\leq1$. If a fraction $f_{\rm host}$ of MBHs host one such
target, the all-sky observer-frame rate is
\begin{align}
 \frac{d\lambda_{{\rm obs},\pm}}{dz}
 &=\frac{f_{\rm host}}{1+z}\frac{dV_c}{dz}
   \int_6^8\Phi_\pm(M_\bullet)\bar\lambda(M_\bullet)\,d\log(M_\bullet/M_\odot),\\
 \lambda_{{\rm obs},\pm}(<z)
 &=\int_0^z\frac{d\lambda_{{\rm obs},\pm}}{dz'}\,dz' .
 \label{eq:sm_cos_rate}
\end{align}
Here $dV_c/dz=4\pi cD_c^2/H(z)$ is the full-sky comoving volume
element and $(1+z)^{-1}$ converts source time to observer time. We use
$D_c(z)=c\int_0^z dz'/H(z')$ in a flat $\Lambda$CDM cosmology with
$H_0=67.4\,\mathrm{km\,s^{-1}\,Mpc^{-1}}$, $\Omega_m=0.315$, and
$\Omega_\Lambda=0.685$~\cite{Planck2018}.

\begin{figure}[t]
 \centering
 \includegraphics[width=\linewidth]{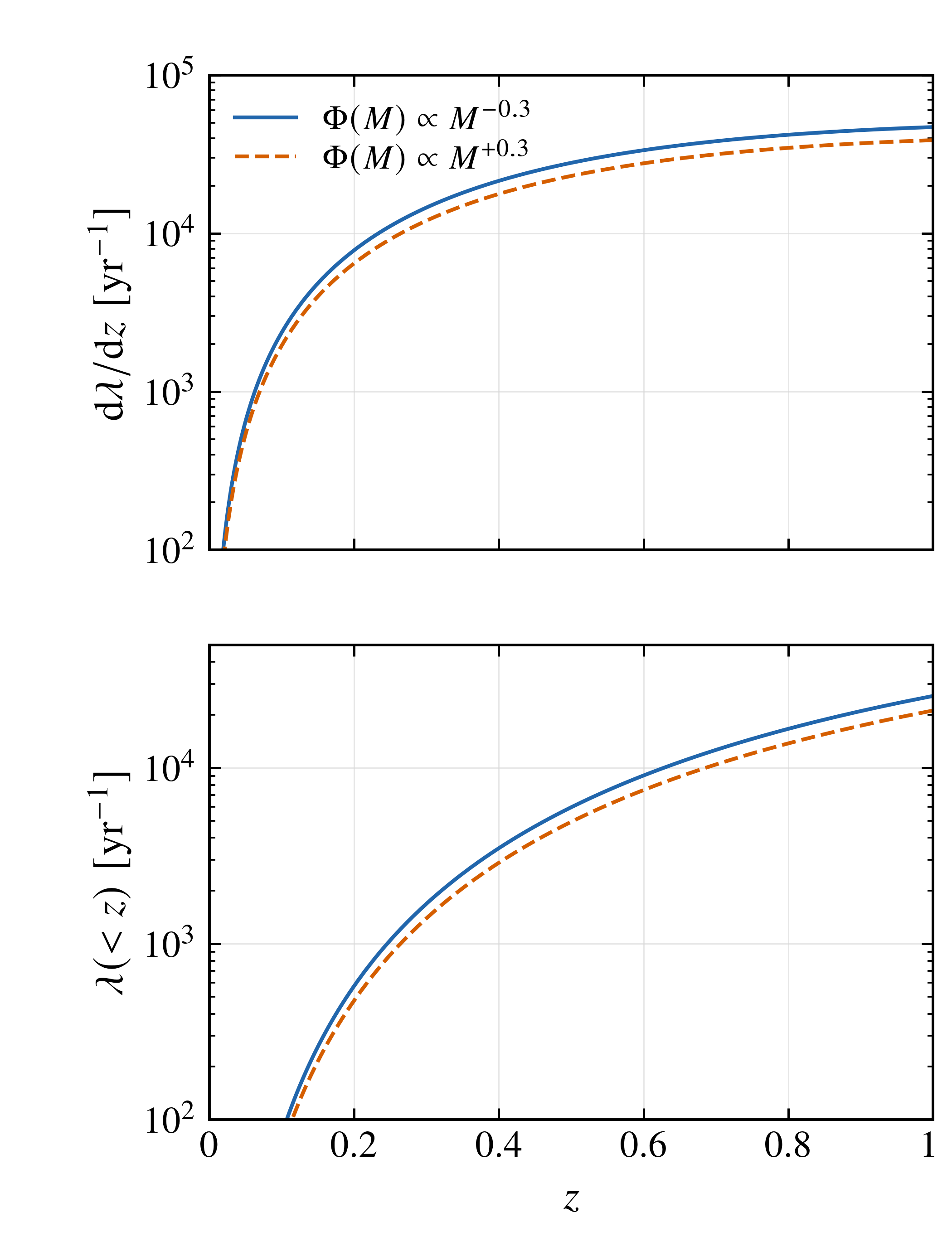}
 \caption{All-sky observer-frame collision rates for one target per
 MBH ($f_{\rm host}=1$). The upper panel gives the differential rate
 and the lower panel gives the cumulative annual rate from sources
 within $z$. The blue solid and orange dashed curves use $\Phi_-$ and
 $\Phi_+$, respectively.}
 \label{fig:sm_redshift}
\end{figure}

Figure~\ref{fig:sm_redshift} gives the normalization for one target per
MBH ($f_{\rm host}=1$). The cumulative rates within $z=1$ are
$2.56\times10^4$ and $2.12\times10^4\,\mathrm{yr}^{-1}$ for $\Phi_-$
and $\Phi_+$, respectively. For the main-text estimate,
$f_{\rm host}=0.1$ gives $2.56\times10^3$ and
$2.12\times10^3\,\mathrm{yr}^{-1}$. The corresponding rates within
$z=0.1$ are $8.29$ and $6.85\,\mathrm{yr}^{-1}$. These rates count
physical contacts before observational selection.

\section{Detailed Collision Simulations}
\label{app:detailed_collision_simulations}

\subsection{Stellar models and encounter geometry}

We follow isolated encounters in the center-of-mass frame with
AthenaK~\cite{Stone2026,2026arXiv260937859J}. The two stars have equal
mass $M_\star$ and radius $R_\star$. Their centers are initially separated
by $4R_\star$ along the collision axis ($x$) and by $b$ in the transverse
direction. The impact-parameter sequence uses nonrotating $n=3/2$
Lane--Emden polytropes with $M_\star=M_\odot$, $R_\star=R_\odot$, and
$\widetilde b=b/(2R_\star)\in\{0,0.1,0.3,0.5,0.7\}$.
After relaxing the stars under self-gravity, we give them opposing
velocities of $0.1c$ along $x$. We measure time $t$ from this launch.
The Newtonian relative speed is $0.2c$, and the total kinetic energy
of the pair is $M_\star(0.1c)^2=1.79\times10^{52}
(M_\star/M_\odot)\,\mathrm{erg}$. We evolve the self-gravity of the
stars and ejecta in a local domain centered on the encounter.

For the head-on mass sequence we evolve nonrotating pre-main-sequence stars with
MESA r26.4.1~\cite{Paxton2011,Jermyn2023}. We adopt initial mass fractions
$(X,Y,Z)=(0.70,0.28,0.02)$, the GS98 metal mixture, and a mixing-length
parameter of 2.0. We use the \texttt{basic.net} reaction network, set
\texttt{mesh\_delta\_coeff} to 0.5 and \texttt{varcontrol\_target} to
$10^{-4}$, and set mass loss, element diffusion, and convective
overshoot to zero. We identify the zero-age main sequence (ZAMS) with
the first timestep at which the nuclear luminosity reaches or exceeds
99\% of the photospheric luminosity. We interpolate the MESA density
and total pressure linearly in radius and extend the endpoint values
to the center and photosphere. Each model retains its native
photospheric radius (Table~\ref{tab:dc_zams}). Before mapping to the
Cartesian grid, the integrated masses differ from the MESA masses by
at most $2.2\times10^{-5}$ in fractional terms. For both stellar sequences
we initialize the gas temperature $T_g$ by solving
\begin{equation}
 p_\star(r)=p_g[\rho(r),T_g(r)]+\tfrac13 a_{\rm rad}T_g^4(r)
\end{equation}
for the prescribed pressure profile $p_\star$, and set the initial
radiation energy density inside the star to $a_{\rm rad}T_g^4$.
Here $a_{\rm rad}$ is the radiation constant. The collision equation
of state supplies the gas internal energy. Before launch, we relax the
solar polytropes for $0.80\,t_{\rm dyn}$ and the ZAMS models for
$0.080\,t_{\rm dyn}$ to ensure the stellar profiles reach quasi-steady state,
where $t_{\rm dyn}\equiv[R_\star^3/(GM_\star)]^{1/2}$
is the dynamical time of each star. During relaxation we damp stellar
velocities with an e-folding time of $0.040\,t_{\rm dyn}$ in both sequences.

\subsection{Hydrodynamics and radiation transport}

The initial Cartesian domain spans $[-8R_\star,8R_\star]$ along each
axis. Five levels of AMR refinement above a $32^3$ root grid
give a finest cell width of $R_\star/64$, or 128 cells across a stellar
diameter. Each meshblock contains $16^3$ active cells. We advect two
passive mass fractions to track the individual stars. When tagged gas
reaches 80\% of the distance to a box face, we double the box in each
dimension and interpolate the fields onto the new mesh. After the first
expansion, refinement follows material at outward radial optical depths
of approximately 1--10.
The physical cell widths and the intervals
between snapshots increase as the ejecta expand. Each run follows
the flow for 71.62 days after launch.

The ambient gas starts at $10^{-18}\,\mathrm{g\,cm^{-3}}$ and
$10^4\,\mathrm{K}$, with radiation at the numerical floor. We impose
outflow-only gas boundaries and zero incident radiation intensity. We solve
$\nabla^2\Phi_{\rm sg}=4\pi G\rho_{\rm grav}$ with geometric multigrid
and isolated multipole boundary conditions. The gravitational force
density in the momentum equation is
$\boldsymbol f_g=-\rho_{\rm grav}\nabla\Phi_{\rm sg}$. Here
$\rho_{\rm grav}$ equals the gas density above a cutoff and vanishes
below it. The cutoff is $7.38\times10^{-12}\,\mathrm{g\,cm^{-3}}$
for the polytropes and $10^{-18}\,\mathrm{g\,cm^{-3}}$ for the ZAMS
models. The finite-volume update uses an HLLE gas solver during
relaxation and HLLC during the collision. In the latter phase we use
bounded piecewise parabolic reconstruction for density and piecewise
linear reconstruction for velocity, specific thermal energy, and
tracers. We use second-order Runge--Kutta (RK2) transport, a Courant (CFL) number 
of 0.3, and five levels of local adaptive time (LAT) stepping. A dual-energy update
recovers the internal energy in kinetic-energy-dominated gas. We impose
a gas-temperature floor of $100\,\mathrm{K}$ and a gas-speed cap of $0.3c$.

\begin{figure*}[t]
 \centering
 \includegraphics[width=0.97\textwidth]{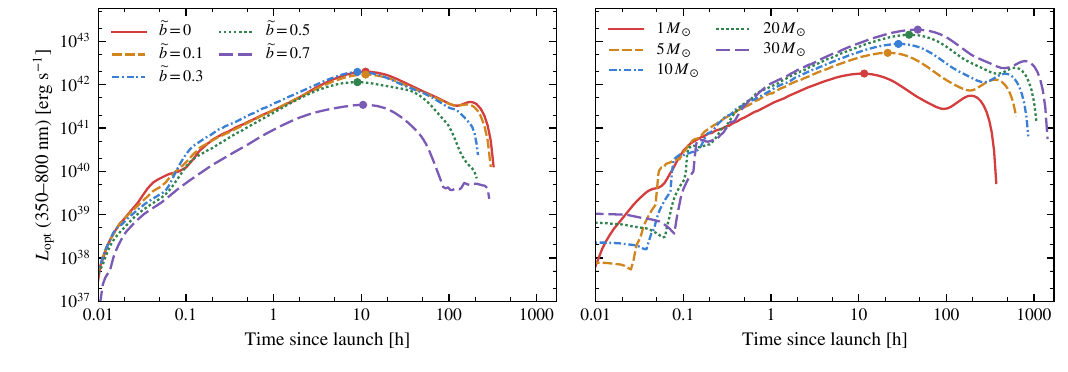}
 \caption{Direct 350--800\,nm luminosity of the cooling ejecta. The
 left panel compares offsets for pairs of solar polytropes. The right
 panel compares stellar masses for head-on ZAMS collisions. Time is
 measured from launch. Dots mark the brightest saved sample of each curve.}
 \label{fig:dc_optical}
\end{figure*}

The gas equation of state uses a uniform H/He mixture with
$(X,Y)=(0.70,0.30)$ and includes dissociation, ionization, dense-fluid
effects, and electron--positron pairs~\cite{Tomida2013,Chabrier2021,Timmes2000}.
The grey transport opacity per unit mass, $\chi_R$, includes electron
scattering with the local ionization fraction and free--free opacity
with a hot-plasma normalization $Z=0.02$. We approximate the absorption
opacity for thermal exchange by $\kappa_a=\chi_R-\kappa_s$, where
$\kappa_s$ is the scattering opacity. During the collision the
microphysical energy sources are matter--radiation exchange and thermal
neutrino cooling. In the gas rest frame (subscript 0), the radiation
four-force on the gas is
\begin{align}
 G_0^0 &= \rho\kappa_a(\mathsf E_{r,0}-a_{\rm rad}T_g^4), \nonumber\\
 \boldsymbol G_0 &= (\rho\chi_R/c)\,\mathsf F_{r,0}.
\end{align}
We use local Lorentz transformations for the moments and four-force,
apply the exchange implicitly with opposite signs for gas and radiation,
and evolve the lab-frame moments at the physical speed of light.
The energy-loss term $\mathcal L_\nu$ includes freely escaping neutrinos
from pair annihilation, photoneutrino production, plasmon decay, and
bremsstrahlung~\cite{Itoh1996}.

To close the radiation pressure tensor $\mathsf P_r$, we solve an
auxiliary stationary transfer problem on straight rays through the
instantaneous gas distribution. With isotropic coherent scattering,
the auxiliary intensity $I_d$ in direction $\hat{\boldsymbol n}$ obeys
\begin{align}
 \hat{\boldsymbol n}\cdot\nabla I_d+\rho\chi_R I_d
 &=\frac{\rho c}{4\pi}
    \big(\kappa_a a_{\rm rad}T_g^4+\kappa_s E_d\big),\nonumber\\
 E_d&=\frac{1}{c}\int I_d\,\mathrm{d}\Omega .
\end{align}
Let $\mathsf q_d$ be the angular covariance of the ray directions,
weighted by $I_d$ and normalized to unit trace. With reduced flux
$\boldsymbol f=\mathsf F_r/(c\mathsf E_r)$, the closure is
\begin{equation}
 \frac{\mathsf P_r}{\mathsf E_r}
 =(1-w)\mathsf D_{\rm M1}
 +w\big[\boldsymbol f\boldsymbol f+(1-|\boldsymbol f|^2)\mathsf q_d\big].
\end{equation}
Here $\mathsf D_{\rm M1}$ is the M1 Eddington tensor~\cite{Levermore1984}
and $w=\exp(-\tau_{\rm bin})$ depends on the transport depth across a
source-grid cell. We use M1 wherever the auxiliary angular distribution
is unavailable. The spherical source grid has $96\times64\times32$
cells in radius, polar angle, and azimuth, with 288 ray directions per
target. We rebuild the closure at the first synchronized step at least
five cycles after the preceding update. We take up to four $\Lambda$
iterations per update and retain the final iterate even when its median
fractional change in radiation energy exceeds 1\%.

\subsection{From grey moments to band luminosities}

At each snapshot we integrate the outward transport optical depth through
the native mesh along rays from the collision origin. For direction
$\Omega$ and distance $r_{\rm box}(\Omega)$ to the outer boundary,
\begin{align}
 \tau_r(r,\Omega) &= \int_r^{r_{\rm box}(\Omega)}\rho\chi_R\,\mathrm{d}s,
 \nonumber\\
 \bar\tau(r) &= \frac{1}{4\pi}\int \tau_r(r,\Omega)\,\mathrm{d}\Omega .
\end{align}
We select the sphere of radius $R_{\rm ph}$ satisfying
$\bar\tau(R_{\rm ph})=1$. The depth on that sphere can vary strongly
with direction. For a band $B$ with frequency limits $\nu_1,\nu_2$,
we assign each surface element a Planck fraction at its local radiation
temperature $T_r=(\mathsf E_r/a_{\rm rad})^{1/4}$,
\begin{align}
 f_B(T_r) &=
 \frac{\int_{\nu_1}^{\nu_2} B_\nu(T_r)\,\mathrm{d}\nu}
      {\int_0^\infty B_\nu(T_r)\,\mathrm{d}\nu},\nonumber\\
 L_B(t) &= R_{\rm ph}^2\int f_B(T_r)\,
       \mathsf F_r\cdot\hat{\boldsymbol r}\,\mathrm{d}\Omega .
 \label{eq:dc_band}
\end{align}
The integral retains the sign of the radial flux. Setting $f_B=1$
gives the bolometric luminosity $L_{\bar\tau=1}$. We use unit energy
response over 350--800\,nm for the optical band and 0.5--10\,keV for
X-rays. We refine the equal-solid-angle quadrature until the full set
and an alternating half-set agree within 2\% for the bolometric flux
and within $0.02|L_B|+10^{25}\,\mathrm{erg\,s^{-1}}$ for each band.

\subsection{Direct optical emission and X-ray durations}

The direct optical light curves peak hours after the X-ray flash
(Fig.~\ref{fig:dc_optical}). For the five polytropic encounters in
order of increasing $\widetilde b$, the optical maxima are
$(2.01,1.72,1.98,1.15,0.345)\times10^{42}\,\mathrm{erg\,s^{-1}}$.
Their peak epochs range from 9.05 to 11.22\,h after launch. The optical
maxima for the head-on and $\widetilde b=0.3$ cases agree within 2\%,
suggesting the stars are fully shocked.
For the fiducial head-on collision of two $30\,M_\odot$ stars,
the bolometric and X-ray peaks coincide at $507\,\mathrm{s}$, when
the X-ray band carries 98\% of the bolometric luminosity.
By the direct optical peak at $47.3\,\mathrm{h}$,
the bolometric luminosity has fallen from $4.38\times10^{48}$ to
$6.15\times10^{43}\,\mathrm{erg\,s^{-1}}$, and the optical band carries
31\% of the escaping flux.

\begin{table}[t]
 \caption{Head-on ZAMS collisions. Mass and radius refer to one star.
 $L_{\rm opt,pk}$ and $L_{\rm X,pk}$ are the brightest saved samples
 in units of $10^{42}$ and $10^{48}\,\mathrm{erg\,s^{-1}}$,
 respectively. $\Delta t_{\rm opt}$ is the rise time from contact to
 optical maximum. $\Delta t_{\rm X}$ is the time above
 $10^{45}\,\mathrm{erg\,s^{-1}}$ in the X-ray band.}
 \label{tab:dc_zams}
 \begin{ruledtabular}
 \begin{tabular}{rrrrrr}
 $M_\star/M_\odot$ & $R_\star/R_\odot$ & $L_{\rm opt,pk}$ &
 $\Delta t_{\rm opt}$ & $L_{\rm X,pk}$ & $\Delta t_{\rm X}$ \\
 & & & (h) & & (s) \\
 \hline
  1 & 0.893 &  1.82 & 11.6 & 0.00558 &  384 \\
  5 & 2.707 &  5.48 & 21.4 & 2.13    & 1373 \\
 10 & 3.944 &  8.80 & 28.4 & 2.67    & 1823 \\
 20 & 5.799 & 14.4  & 37.6 & 3.71    & 2731 \\
 30 & 7.281 & 18.8  & 47.2 & 4.29    & 3391
 \end{tabular}
 \end{ruledtabular}
\end{table}

For Table~\ref{tab:dc_zams} we estimate contact by linearly
interpolating the first crossing of $2R_\star$ by the separation of
the two tracer-mass centroids. We also interpolate $L_{\rm X}$
linearly in time at a chosen luminosity threshold and sum the intervals
above it. Each ZAMS pair has one interval above
$10^{45}\,\mathrm{erg\,s^{-1}}$, lasting from 384 to $3391\,\mathrm{s}$
across the sequence. Lowering the threshold to
$10^{43}\,\mathrm{erg\,s^{-1}}$ includes more of the fading emission
and gives durations from 0.32 to 1.82\,h.

The stellar envelopes help explain the difference between the prompt
flashes. In the initial ZAMS profiles the mass column exterior to
$0.8R_\star$ is $7.2\times10^8\,\mathrm{g\,cm^{-2}}$ for
$1\,M_\odot$ and $(1.5$--$2.3)\times10^7\,\mathrm{g\,cm^{-2}}$ for
$5$--$30\,M_\odot$. At $ct/R_\star\simeq30$, the median radial depth
from $4R_\star$ to the box boundary in the two cones
$|\cos\theta_x|>0.95$ is 19.5 for $1\,M_\odot$ and
$1.0\times10^{-2}$ for $5\,M_\odot$. Here $\theta_x$ is measured from
the collision axis. The cones cover 5\% of the sky and carry about
7\% and 93\% of the positive outward X-ray power on the extraction
sphere, respectively. The thicker polar column in the solar-mass
model is consistent with its later X-ray maximum.

For the reprocessed component of the fiducial collision we drive the
external cloud with the measured $L_{\bar\tau=1}(t)$.
We set the stored energy to zero at launch, interpolate the incident
power linearly between snapshots, and adopt $f_{\rm rep}=0.4$,
$n_e=3\times10^{10}\,\mathrm{cm^{-3}}$, and
$R_{\rm rep}=10^3GM_\bullet/c^2$ for $M_\bullet=10^7M_\odot$.
These values give $t_{\rm esc}=16.8$ days. We compute the optical
fraction of the escaping cloud luminosity $E/t_{\rm esc}$ at the
temperature defined by
$4\pi R_{\rm rep}^2\sigma_{\rm B}T_{\rm eff}^4=E/t_{\rm esc}$.
The cloud's optical luminosity peaks at
$3.29\times10^{43}\,\mathrm{erg\,s^{-1}}$ about 11\,h after launch.
For a clear line of sight we add the cloud and direct fireball
luminosities. After the last measured fireball sample, the cloud model
evolves its stored energy by exponential decay on $t_{\rm esc}$.

\def\bibsection{\section*{References}}
\bibliography{reference.bib}

\end{document}